\documentclass[twocolumn]{aastex631}

\usepackage{multirow}  
\usepackage{breqn}
\usepackage{gensymb}

\submitjournal{ApJ}

\begin{document}

\title{{Foreshortening of Polarizer Angles} in Wide Field Polarizing Imagers}

\correspondingauthor{Ritesh Patel}
\email{ritesh.patel@swri.org}

\author[0000-0001-8504-2725]{Ritesh Patel}
\affiliation{Southwest Research Institute,
1301 Walnut Street, Suite 400,
Boulder, CO 80302, USA}

\author[0000-0003-3410-7650]{J. Marcus Hughes}
\affiliation{Southwest Research Institute,
1301 Walnut Street, Suite 400,
Boulder, CO 80302, USA}

\author[0000-0002-7164-2786]{C.E. DeForest}
\affiliation{Southwest Research Institute,
1301 Walnut Street, Suite 400,
Boulder, CO 80302, USA}

\author[0000-0002-0494-2025]{Daniel B. Seaton}
\affiliation{Southwest Research Institute,
1301 Walnut Street, Suite 400,
Boulder, CO 80302, USA}

\author[0000-0001-8318-8229]{Chris Lowder}
\affiliation{Southwest Research Institute,
1301 Walnut Street, Suite 400,
Boulder, CO 80302, USA}

\author[0000-0002-0631-2393]{Matthew J. West}
\affiliation{Southwest Research Institute,
1301 Walnut Street, Suite 400,
Boulder, CO 80302, USA}
\affiliation{European Space Agency, European Space Research and Technology Centre, Noordwijk, Netherlands}

\begin{abstract}
The Polarimeter to Unify Corona and Heliosphere (PUNCH) is a very wide field polarizing imager covering a 90$\degree$ diameter region of the sky, stitched together from four individual fields of view (FOVs), three of which are supported by Wide Field Imagers (WFIs) with 40$\degree$ diameter individual FOVs.  The science objective requires precise measurement of linear polarization.  The WFI FOV is sufficiently wide that polarizer angles are foreshortened by projection effects relative to the line of sight of each pixel in the instrument, deviating from their nominal angular positions of -60$\degree$, 0$\degree$, and +60$\degree$ relative to the instrument frame of reference. These angles are foreshortened by the very wide WFI field of view, changing their effective angle across the image plane – an effect that is analogous to similar challenges of polarimetric right/left channel separation in large-format 3D cinema, such as the IMAX format. {Using a simple geometrical approach we estimated the perceived polarizer angle in the WFI FOV will vary by $\pm$2.5$\degree$ for $\pm$60$\degree$ and $\pm$3.5$\degree$ for 0$\degree$ polarizer angles.} We present the effect of this deviation on the polarized brightness to be recorded by WFI, along with how optical distortion will shift the location of foreshortened angles. We also discuss the implications of this effect on upcoming technologies based on polarimetry across in wide fields.

\end{abstract}


\section{Introduction}
    \label{sec:intro}

The white-light solar corona consists of a near-constant F-Corona and evolving K-Corona \citep{vdHulst50}. The K-corona is predominantly observed through Thomson scattering, resulting in its polarization being oriented at a right angle to the plane formed by the observer, the Sun, and the point of scattering \citep{Billings1966}. The direction of the K-corona's polarization is therefore aligned to the tangent to the solar limb. Measurement of polarization in the corona and heliosphere has been employed to estimate the coronal electron density \citep{vdHulst50, Saito1977SoPh}, for the tomographic reconstruction of 3D electron density in the corona {\citep{Davila1994ApJ, Zidowitz1999JGR, Frazin2002ApJ, Kramar2009SoPh, Morgan2015ApJS}}, to measure 3D characteristics of coronal mass ejections \citep[CMEs;][]{Moran2004Sci, Mierla2009SoPh, DeForest2013a, Susino2014ApJ}, and to determine the chirality of CMEs \citep{DeForest2017ApJ}. Based on an analogy between linear polarization and human color perception \citep{DeForest_etal2022}, \citet{Patel2023RNAAS} used the 2023 total solar eclipse observations of the middle corona \citep{West2023} from the Citizen CATE 2024 project \citep[][]{Kovac2025arXiv} to set hue encoding of polarization angle \citep[e.g.,][]{hoeksema2014helioseismic} on a rigorous analytical footing.

Coronal polarization has been previously measured by many space-based coronagraphs including Skylab \citep{SKYLAB}, the Solar and Heliospheric Observatory Large Angle and Spectrometric Coronagraph Experiment  \citep[SOHO/LASCO;][]{Brueckner95}, Sun Earth Connection Coronal and Heliospheric Investigation \citep[SECCHI;][]{Howard2008SSRv} of Solar Terrestrial Relations Observatory and Solar Orbiter Metis \citep{Antonucci2020A&A}. Recently launched missions including Association of Spacecraft for Polarimetric and Imaging Investigation of the Corona of the Sun \citep[ASPIICS;][]{Shestov2021A&A, Zhukov2025arXiv} onboard Project for Onboard Autonomy-3, Coronal Diagnostic Experiment on International Space Station \citep{Cho2017JKAS, Newmark2020AGUFMSH}, and Polarimeter to Unify Corona and Heliosphere \citep[PUNCH;][]{DeForest2022_punch, DeForest2025arXiv} also include white-light imaging polarimeters to probe the corona and heliosphere. The largest field of view (FOV) among the previously flown polarimeters was $\pm$8$\degree$ of LASCO/C3.

The PUNCH Wide Field Imager {\citep[WFI;][]{laurent2026_wfi}} has a truncated square field of view (FOV) with a maximum diagonal over 50$\degree$. Such a wide FOV introduces new challenges for polarimetry. In particular, objects located at wider incident angles view the instrument polarizer angles foreshortened from their actual angles relative to the instrument boresight, an effect that has been negligible in prior instruments with smaller FOVs.

Previously, the effect of wide viewing angles has been explored for the case of liquid crystal displays \citep[LCDs;][]{Yang06} as {LCDs} also leverage the phenomenon of polarization. However, \citeauthor{Yang06} found that even though viewing angles up to 140$\degree$ could be achieved, they face an issue of light leakage from a weak/spoiled transmission null in the LCD. To overcome such challenges, \citet{Ishinabe2002JaJAP} introduced a new type of wide viewing angle polarizer. {Early LCD displays were therefore the subject of much research on the topic of wide-field polarization measurements, with a nice early review by  \citet{Hong2006}. More recently}, photometric and colorimetric properties of polarization were explored by \citet{LeDeun2019EffectOV} in multi-viewer projector based wide screens. Such systems have been improved over the years with the development of materials optimized for large viewing angles.  While LCDs are a different system from polarimetric coronagraphs, there are strong analogies that can be drawn between the two systems' responses to optical projection of the polarizer angle across a wide FOV.

In this article we describe the effect of large viewing angle with respect to wide field polarizer-based imagers in context of {heliospheric imaging, particularly with PUNCH}. In Section~\ref{sec:imax} we introduce the {foreshortening effect on polarizer angles due to wide FOV}, analyze the effects of projecting curved celestial sphere on a plane (detector) and discuss its impact for deriving polarized brightness in PUNCH. We finally summarize the work in Section~\ref{sec-conclusion}.


\section{{foreshortening effect}}
    \label{sec:imax}

\subsection{Overview}
    \label{ss:overview}

\begin{figure*}[!ht]
    \centering
    \includegraphics[width=0.9\linewidth]{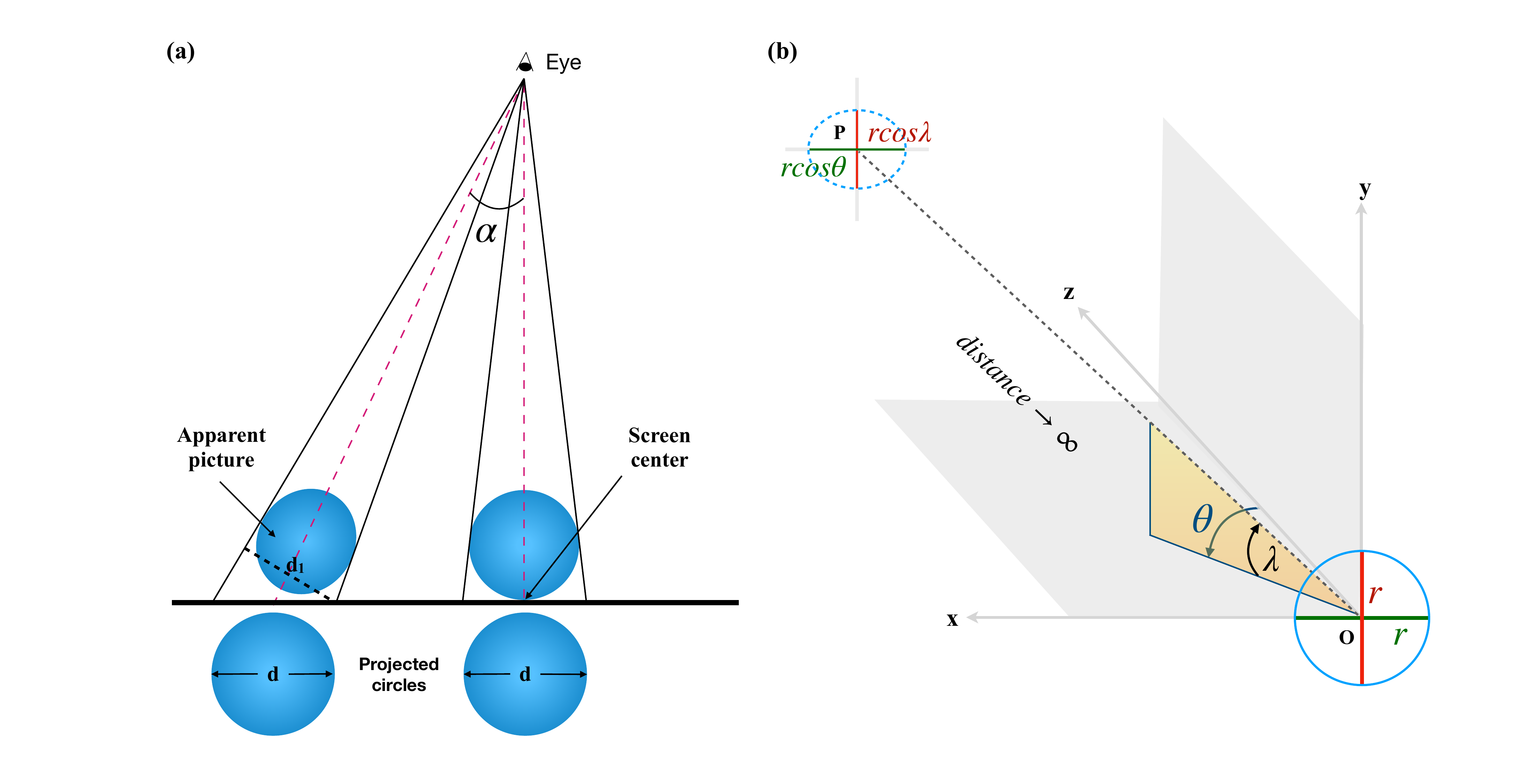}
    \includegraphics[width=0.9\linewidth]{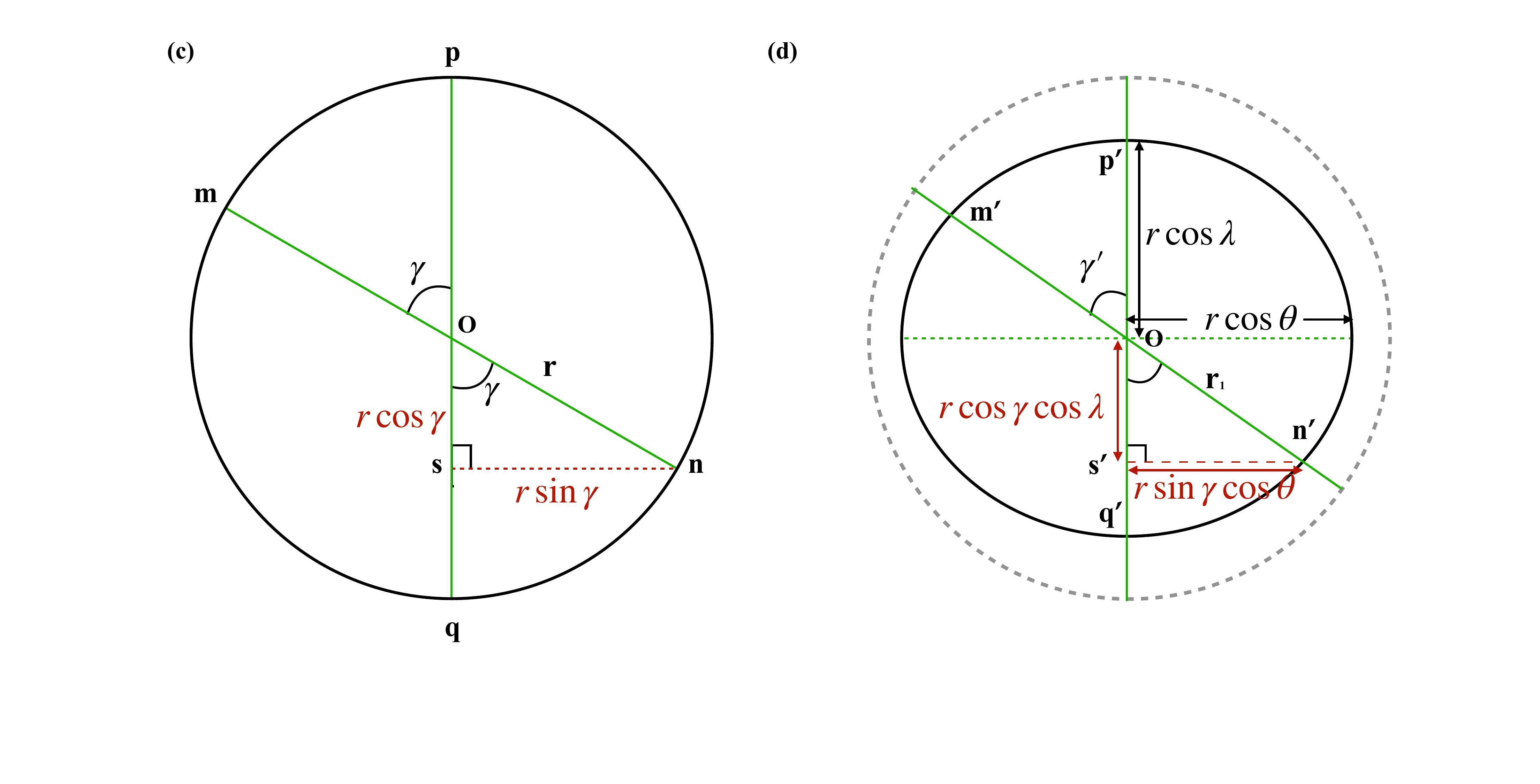}
    \caption{We understand polarizer {foreshortening effect}s in the WFI instrument, by
    analogy to the terrestrial system of wide-field cinema.  (a) Image distortion in a {wide-screen} cinema flat screen when the viewing angle changes by $\alpha\degree$ from the center. (b) Analogy to {a cinema} screen where two orthogonal lines {within the circle} at point O with dimension \textit{r} are viewed from point \textit{P} located at {longitude and latitude} of $\lambda$ and $\theta$ respectively and {and appear as ellipse due to foreshortening}. {The yellow triangular plane shows the angular separation of point P with respect to the reference system in 3D.} (c) and (d) represents the distortion of a circle with radius r to an ellipse when viewed from a point at distance $\to\infty$ similar to (b) located at $\lambda$ and $\theta$.}
    \label{fig:imax_bg}
\end{figure*}    

\begin{figure*}
    \centering
    \includegraphics[width=1.035\linewidth]{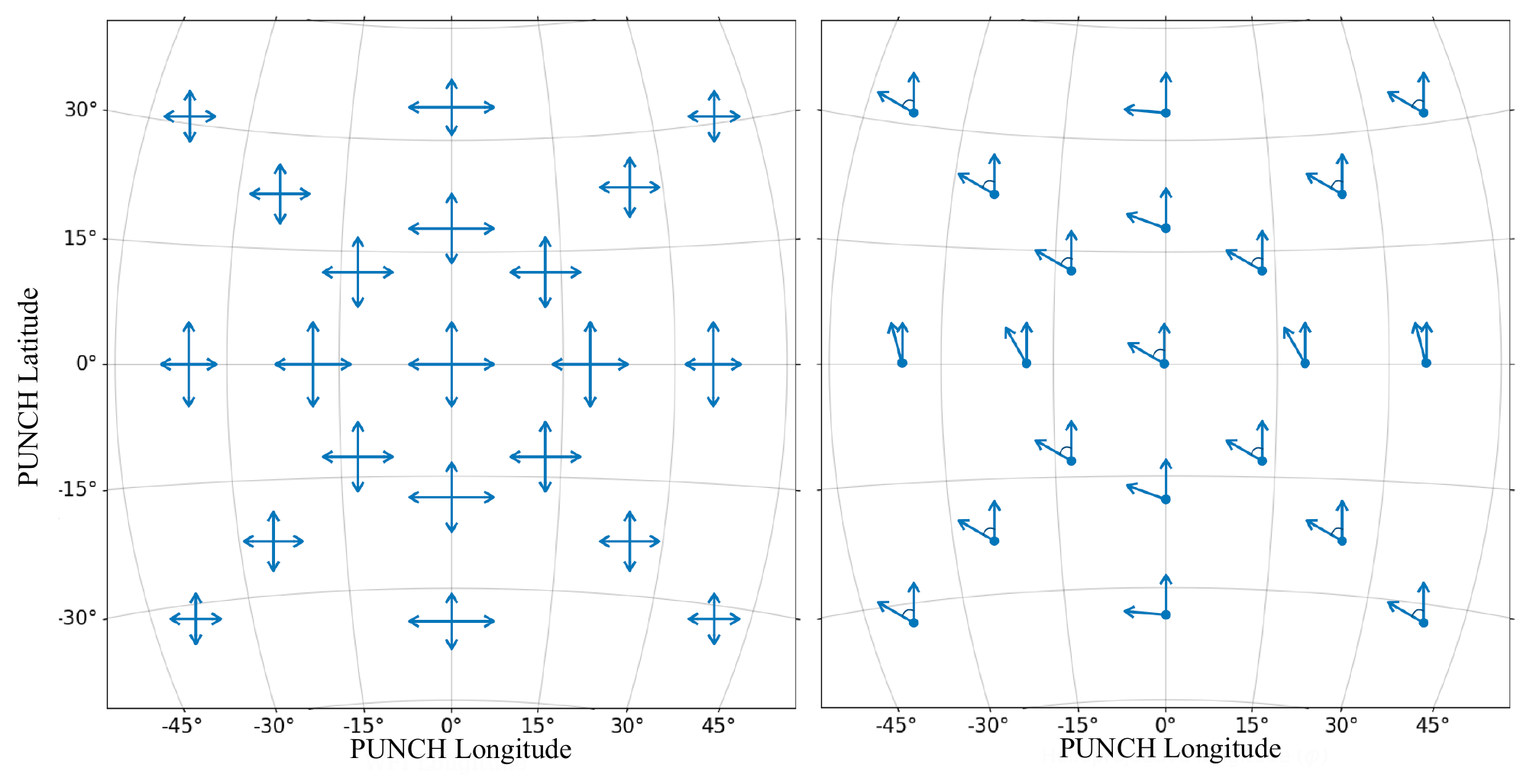}
    \caption{{Plane-parallel projection versus spherical sky geometry. Left to right: apparent lengths along two orthogonal directions and the resulting effective polarizer angles under a plane-parallel projection at the detector plane, illustrating the {foreshortening effect} for a wide field of view. The background grid shows the curved celestial sphere projected using the azimuthal–equidistant (ARC) projection adopted for PUNCH data products. For clarity, the distortions are exaggerated in both panels. When applied to the vector components of an optical polarizer, this projection-induced distortion leads to systematic, field-dependent rotations of the inferred polarization direction across the wide field of view.}}
    \label{fig:imax_gen}
\end{figure*} 

{The projection of a flat optical polarizer into the wide FOV of the WFI instrument is directly analogous to polarization projection effects in the context of 3D cinema viewing, and we therefore explore that system to build intuition and relate the problem to the existing literature in that field}. Figure~\ref{fig:imax_bg}a illustrates how a {small} circle{, displayed on a cinema screen}, will be projected when viewed at center and at a horizontal angle $\alpha$ from the center. In the latter case, the circle with diameter \emph{d} {appears} as an ellipse with {apparent minor axis length} \emph{d$_1$},(see Figure~\ref{fig:imax_bg}a), producing a distortion factor (n) given by:

\begin{equation}
\label{eq-distort}
    n = \frac{d}{d_1} = \frac{1}{\cos{\alpha}}.
\end{equation}


In {wide-screen 3D cinema systems}, polarizer foreshortening is a concern because polarized glasses are used to separate the right and left visual channels for stereoscopy, using {either} orthogonal linear polarizing filters {or opposed circular filters}. In the context of 3D cinema, {projection effects in the polarization signal itself induce} crosstalk between left- and right-eye channels from the wide field {because all spatial dimensions, including those of physical linear polarizers themselves, are subject to the projection/distortion effect described in Equation \ref{eq-distort}}.

A schematic highlighting the geometrical distortion of two foreshortened lines is shown in Figure~\ref{fig:imax_bg}b, where line segments of length $r$ located at a source point $O$, and the perceived distances due to the foreshortening observed by an observer positioned at point $P$, a position where a plane-parallel projection\footnote{The source is assumed to be sufficiently distant that incoming rays are treated as parallel.} is assumed.  $P$ is effectively located at a very large distance ($\rightarrow$ infinity), and the angular coordinates of the projected lines are at $\theta$ and $\lambda$ with respect to the reference position {along the azimuthal and elevation direction respectively}.  

{To understand the angular effects on short line segments, such as physical polarizing elements in a linear polarizer, we resolve the foreshortening into vertical and horizontal components.} Using Equation~\ref{eq-distort}, the projected lengths of {short vertical and horizontal} line { segments} ($r$) will be $r \cos{\lambda}$ and $r \sin{\theta}$, respectively. Similarly, a circular geometry, represented in Figure~\ref{fig:imax_bg}c, treated using similar projections, will be distorted to an ellipse, as shown in panel\,d. {Since a short line segment may be considered as a radius of a small circle, characterizing
this ellipse is equivalent to characterizing the effect
of foreshortening at a single point, on the entire
ensemble of possible 
polarizer orientations at that point.}

{Foreshortening affects} the angle $\gamma$ subtended by an arc in the circle shown in Figure~\ref{fig:imax_bg}c, measured counterclockwise from the north-south vertical diameter ($p-q$) in the circle. A projected arc with an angle $\gamma^{\prime}$ is shown in panel d.  If we consider the right-angled triangle with corners \emph{n}, \emph{s}, and \emph{o}, the sides \emph{n-s} (opposite $\gamma$) and \emph{o-s} (adjacent to $\gamma$) can be represented as sine and cosine components of the hypotenuse, \emph{n-o}, which is also equal to radius of the circle, \emph{r}. In the foreshortened projection of this circle (Figure~\ref{fig:imax_bg}d), the lengths of the line segments within the circle are modified in a way similar to the vertical and horizontal lines presented in Figure~\ref{fig:imax_bg}b, and as a result, the sides of the foreshortened triangle can be represented as,
\begin{equation}
\label{eq-ellipseh}
    n^{\prime}\;s^{\prime} = r\;\sin{\gamma}\;\cos{\theta},
\end{equation}

\begin{equation}
\label{eq-ellipsev}
    o\;s^{\prime} = r\;\cos{\gamma} \cos{\lambda}.
\end{equation}

Using the trigonometric relation in the right triangle $n^{\prime}s^{\prime}o$ and solving for the foreshortened angle $\gamma^{\prime}$:
\begin{equation}
    \label{eq-gamma}
    \gamma^{\prime} = 
    \tan^{-1}\left( \frac{r\sin{\gamma}\cos{\theta}}
    {r\cos{\gamma}\cos{\lambda}}\right)=
    \tan^{-1}\left(\frac{\tan{\gamma} \cos{\theta}}{\cos{\lambda}}\right).
\end{equation}
Equation~\ref{eq-gamma} is a simple way to estimate the foreshortened angle if the location of the observation is known with respect to a plane center.  {In the case of a planar polarizer in front of a wide-field imager such as WFI, the small
polarizing elements in the polarizer are projected onto the plane perpendicular to each 
line of sight and therefore a flat polarizer aligned with the instrument boresight and rotated to angle $\gamma$ relative to the instrument acts, for lines of sight different from the boresight itself, as if it were mounted at an angle $\gamma'$.}

The effect of foreshortening, derived using Equations~\ref{eq-ellipseh} and~\ref{eq-ellipsev}, on two orthogonal axes when observed by a distant observer is shown in the left panel of Figure~\ref{fig:imax_gen}. {The blue vectors illustrate a plane-parallel projection, analogous to a Cartesian grid in which lines of latitude and longitude remain parallel and uniformly spaced, resembling a two-dimensional matrix. The background grid is shown for reference and represents the curved celestial sphere projected using an azimuthal–equidistant (ARC) projection, as adopted in PUNCH data products. This comparison highlights the geometric discrepancy that arises when a locally planar approximation is applied to an intrinsically curved sky.} The length of one or both of the axes get shortened with variation in the viewing angle. Along the lines of $0\degree$ longitude or latitude the vertical or horizontal axis is shortened, respectively. Whereas, along the diagonal direction the shortening is proportional along both axes, maintaining the aspect ratio, but reducing the the overall size of the perceived axes. {That portion of the image is compressed along the radial diagonal but not in the perpendicular direction; this additional effect is not shown in this figure.} {Short vectors that are either radial to the central point, or perpendicular to a radial line (i.e. ``tangential''), remain unchanged.  Short vectors at oblique angles to the radial are shifted slightly in angle due to the shortening.} The right panel of the same figure illustrates the variation in the observed angles derived using Equation~\ref{eq-gamma}. The perceived angles are seen to increase along one of the orthogonal axes (in the vertical direction for the presented example) whereas they shrink along the other axes. 

\subsection{{Effect of Sky Curvature}}
    \label{ss-curvature}

\begin{figure*}
    \centering
    \includegraphics[width=1.035\linewidth]{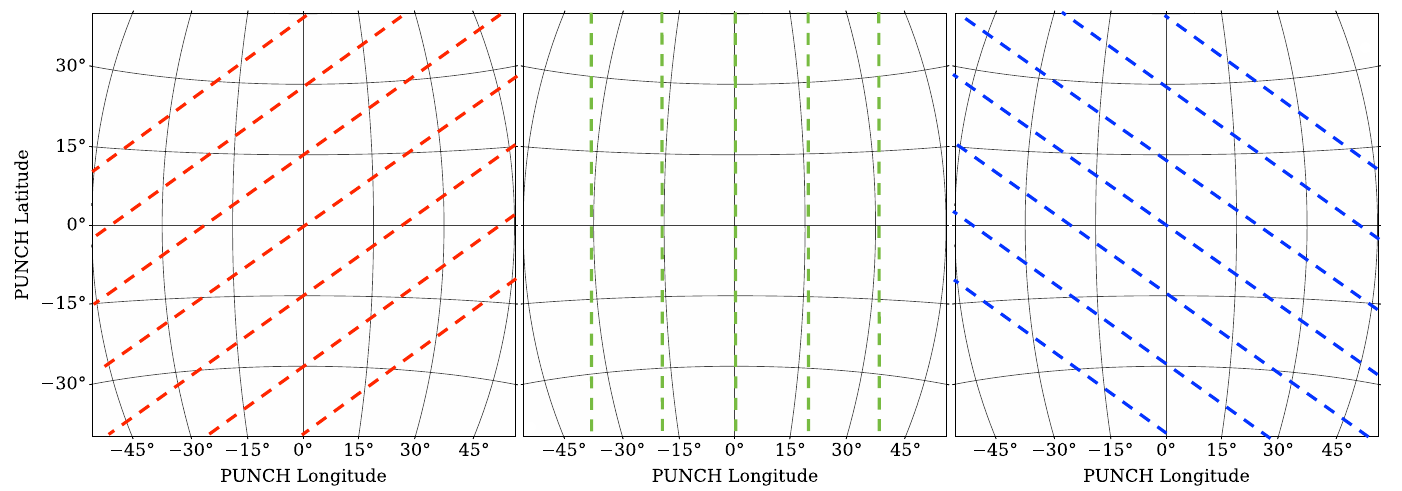}
    \caption{{Effect of sky curvature compared to the detector-plane polarizer orientation. The grid shows the PUNCH FOV in an ARC projection. {\it Left to right:} Polarizer orientations in the detector plane are shown as red, green, and blue dashed lines corresponding to polarizer angles of $-60\degree$, $0\degree$, and $+60\degree$ relative to the vertical direction. The corresponding sky-projected orientations exhibit noticeable deviations, illustrating how detector-plane angles distort when mapped onto the curved sky.}}
    \label{fig:imax_grid}
\end{figure*}     

{When a fixed polarizer orientation in the detector plane is mapped to the celestial sphere through a non-linear projection (e.g., azimuthal perspective, AZP or equidistant, ARC), the apparent orientation of a given direction is distorted, as illustrated in Figure~\ref{fig:imax_grid}. In the detector-plane, the orientation of the polarizers are depicted with dashed lines. Due to the effect of the image projection, solar north deviates significantly from detector vertical at many locations within the FOV. We compute the resulting deviation between the detector-plane polarizer orientation and the corresponding \emph{true-sky} orientation.
\\
For a pixel with helioprojective longitude and latitude $(\theta,\lambda)$ (in radians), we compute its angular distance $\eta$ from the reference direction $(0,0)$ using the {haversine} formula,}
\begin{equation}
a = \sin^{2}\!\left(\frac{\lambda}{2}\right)
  + \cos\lambda \, \sin^{2}\!\left(\frac{\theta}{2}\right),
\end{equation}
\begin{equation}
\eta = 2 \arcsin(\sqrt{a}),
\end{equation}
{which gives the great-circle separation on a unit sphere. This quantity is required because the mapping between detector-plane angles and spherical angles is nonlinear and depends on the local angular distance from the reference point.}


{The detector plane radial direction from image center $(x_{\rm c}, y_{\rm c})$ to an arbitrary pixel at coordinates $(x,y)$ is given by:}
\begin{equation}
\delta = \mathrm{arctan2}\!\left( \frac{-(x - x_{\rm c})} {y - y_{\rm c}} \right),
\end{equation}
{which yields an angle measured from detector $+Y$ in counterclockwise direction. The relative angle between the polarizer axis and the local radial direction in the detector plan is therefore}
\begin{equation}
\beta = \gamma - \delta,
\end{equation}
{specified within the interval $(-\pi,\, \pi]$ to avoid full $2\pi$
redundancies.\\
A line with detector-plane orientation $\beta$ does not retain the same orientation once projected onto the sphere at angular distance $\eta$. The sky-projected orientation $\gamma'$ is given by the classical expression for the projection of a local tangent-plane angle onto a spherical surface such as the notional ``celestial sphere'' at infinite distance:}
\begin{equation}
\gamma' = \arctan\!\left[ \tan(\beta)\,\cos\eta \right].
\label{eq:beta_prime}
\end{equation}
{The deviation between the apparent detector-plane angle and the true-sky orientation is therefore}
\begin{equation}
\Delta\gamma = \gamma' - \beta,
\end{equation}
where
\begin{equation}
-\frac{\pi}{2} \le \Delta\gamma \le +\frac{\pi}{2}.
\end{equation}
%

{Equation~\ref{eq:beta_prime} shows that the distortion increases as a function of angular distance from the FOV center and is proportional to $\cos\eta$---that is, sky orientations become increasingly compressed (relative to planar projections) as $\eta$ grows.}

{For each pixel we compute}
\begin{equation}
\label{eq-imax_map}
{\rm IMAX}(x,y) =
\Delta\gamma(x,y)\,\frac{180}{\pi},
\end{equation}
{yielding a map of polarization angle deviation (in degrees) across the PUNCH FOV, which is shown in Figure~\ref{fig:imax_wfi}.  
This quantity represents the correction required to convert the measured (detector-plane) polarizer angle to the physically relevant sky orientation at any location in the FOV. } 

\subsection{Effect on PUNCH/WFI}
    \label{ss-WFI}


\begin{figure*}
    \centering
    \includegraphics[width=1.035\linewidth]{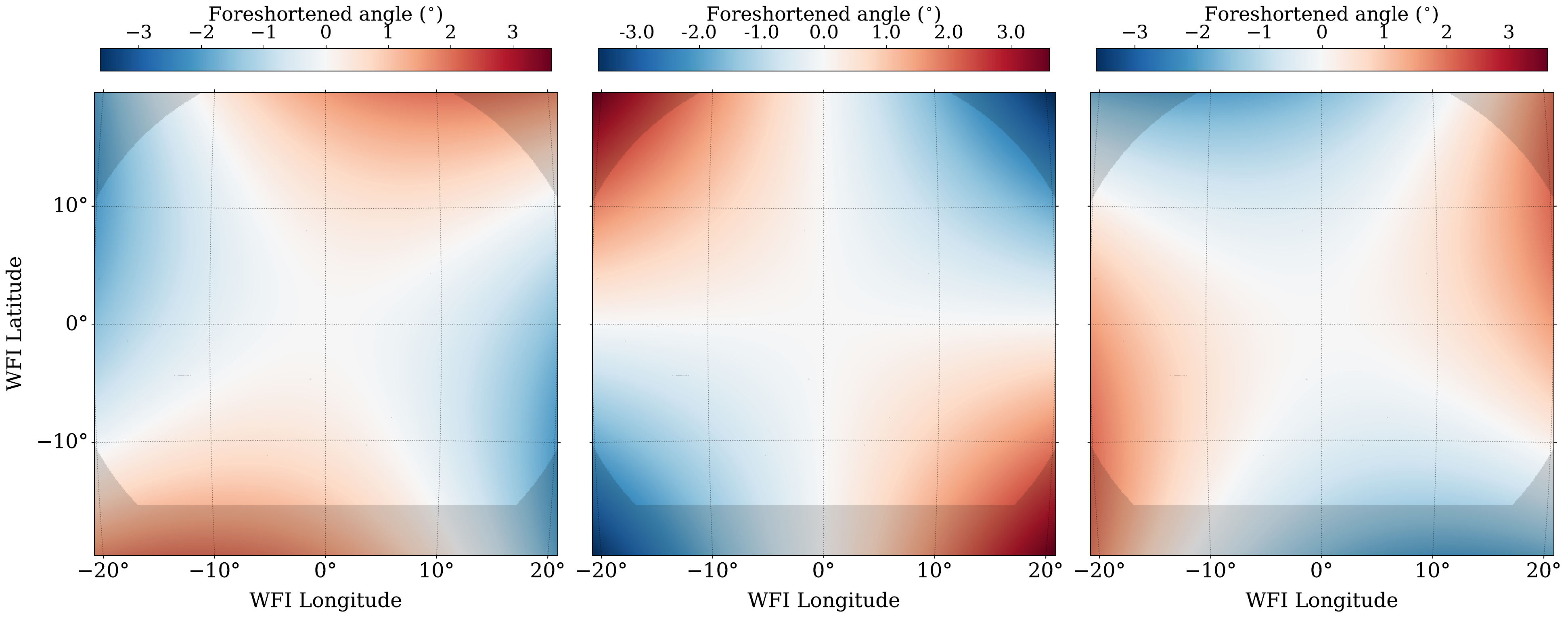}
    \caption{{foreshortening effect} on the three polarizer angles of WFI observed in the instrument frame. {The WFI latitude and longitude correspond to $\lambda$ and $\theta$, described in Section \ref{ss:overview}.} The panels from {\it left} to {\it right} correspond to -60$\degree$, 0$\degree$, and 60$\degree$ respectively and the colors represent the deviation angle {$\Delta\gamma$} measured at a given location in sky. The non-shaded area in the three panels represent the actual useful WFI FOV.}
    \label{fig:imax_wfi}
\end{figure*} 


\begin{figure*}
    \centering
    \includegraphics[width=1\linewidth]{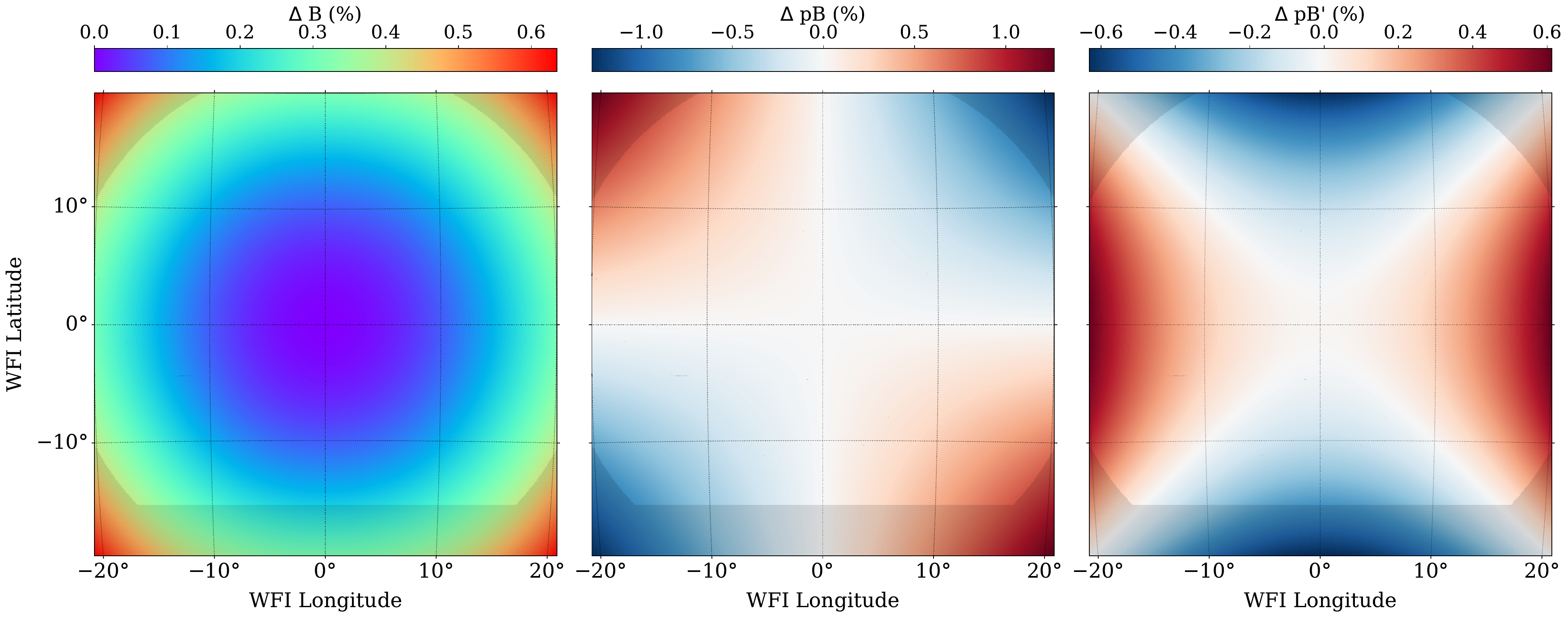}
    \caption{Variation in the polarized brightness in WFI FOV due to {foreshortening effect} when passed through polarizers at angles 0$\degree$ and $\pm$60$\degree$. The colorbar at the top represents the brightness relative to ideal condition.  The variation follows the pattern radially outwards as is expected due to curvature in the FOV projection.}
    \label{fig:imax_fraction}
\end{figure*} 

The PUNCH Wide Field Imagers (WFI) and Near Field Imager (NFI) are equipped with polarizers at angles of -60$\degree$, 0$\degree$, and +60$\degree$, a symmetric three-polarizer measurement and representation system (which we refer to as ``MZP" for ``minus'', ``zero'' and ``plus'', following  DeForest et al. 2022). {The wide FOV of WFI (up to $50\degree$ across) changes the effective angle of each polarizer,
depending on location within the FOV.}  Furthermore, the reference angle used for the MZP system changes with respect to focal plane vertical, across the image plane.  The difference in these angles depends on the angles $\lambda$ and $\theta$, the angles measured across the sky from the center of the image, in the horizontal and vertical directions respectively. An overview of this effect is presented in {Section~\ref{ss:overview}.}


Figure~\ref{fig:imax_wfi} shows three maps of foreshortened polarizer angles across the PUNCH/WFI FOV for the three polarizer angles, $\gamma$ as -60$\degree$, 0$\degree$, and +60$\degree$, respectively. {Over the full square FOV, WFI mask corresponding to the real data is overlayed in gray to show the useful FOV}. The center of the images corresponds to $\theta$ = 0$\degree$ and $\lambda$ = 0$\degree$, and the whole field is defined with respect to the camera coordinates of longitude and latitude respectively {as observed by WFI}. The colors in each panel represent the foreshortened angle (IMAX map from Equation~\ref{eq-imax_map}), which is the difference in perceived angle from the true angle across the FOV. Across the full square field of view, the -60$\degree$ and +60$\degree$ polarizers are shifted up to $\pm$2.5$\degree$ by this effect. For the $\gamma$ = 0$\degree$ polarizer the foreshortening is up to $\approx$3.5$\degree$.

\subsection{Effect on Polarimetric Analysis}
    \label{ss-polarimetric}


{We design a simple experiment using an idealized three-polarizer (MZP) system to assess the impact of the {foreshortening effect} on polarimetric measurements. A linearly polarized input signal with a polarization fraction of 10\% is assumed, with the normalization, similar to expected polarization signal levels from heliosphere. Assuming that this polarization is along the vertical direction such that the intensity recorded by the $0^\circ$ polarizer (Z) is 10\%.
For an ideal instrument, the intensity recorded by each polarizer is given by the squared cosine projection of the incident electric field onto the polarizer transmission axis yielding 2.5\% polarized signal for M and P polarizers. The three measured intensities, $B_M$, $B_Z$ and $B_P$, are combined to estimate the total brightness (B), and polarized brightness (pB) and pB' analogous to Stokes I, Q and U in the solar reference frame using equations 7, 9 and 10 of \cite{DeForest_etal2022}. }

{Using the three maps shown in Figure~\ref{fig:imax_wfi} we generate a cumulative angular perturbation map characterized by adding the three angular deviation maps in quadrature. The resultant map represent the net departure of the perceived polarizer orientations from their ideal configuration. We use the following equations from \cite{DeForest_etal2022} to estimate the fractional errors in B, pB, pB' due to polarizer misalignment:}

\begin{equation}
\Delta B = \sqrt{\frac{2}{3}} \,\sqrt{pB^2 + {pB'}^2}\,\Delta\theta,
\end{equation}

\begin{equation}
\Delta pB = 2\sqrt{\frac{2}{3}} \,pB' \,\Delta\theta,
\end{equation}

\begin{equation}
\Delta pB' = 2\sqrt{\frac{2}{3}} \,pB \,\Delta\theta,
\end{equation}
where angles are expressed in radians.

{Figure~\ref{fig:imax_fraction} shows these fractional difference between the expected and measured B, pB and pB'. We find that the {foreshortening effect} introduces a systematic variation of $\approx$0.5\% from the center to the edge of the WFI FOV.}




\subsection{Effect on Polarization Gamut}
    \label{ss-gamut}

    \begin{figure*}
    \centering
    \includegraphics[width=0.95\linewidth]{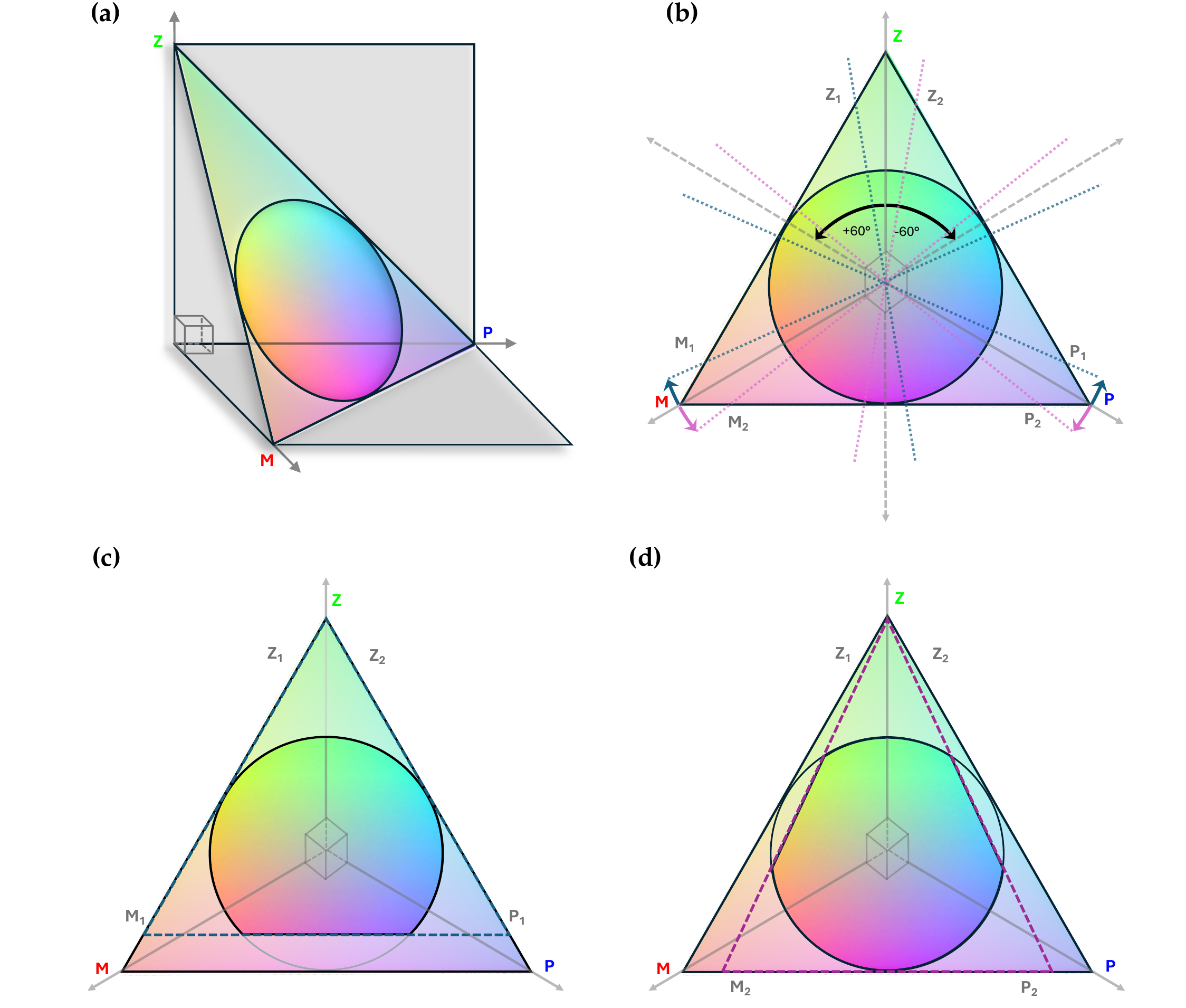}
    \includegraphics[width=0.95\linewidth]{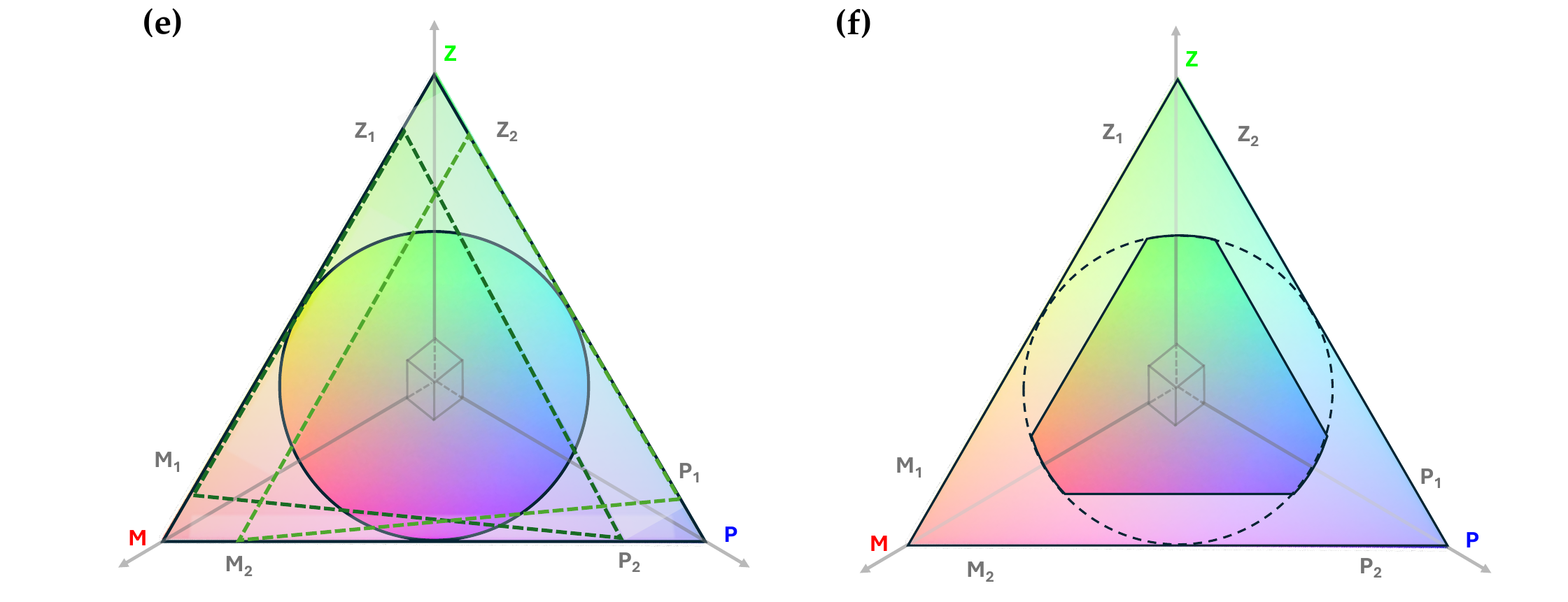}
    \caption{Effect on the polarization gamut for WFI FOV where MZP states are represented analogously to RGB colors. {(a) Reference Cartesian coordinate system for MZP space following Figure 2 of \citet{DeForest_etal2022}; the triangular plane shows the ideal polarization gamut with an inscribed circle indicating the reproducible region. (b) Face-on view of the gamut (note the cube orientation), with gray dotted arrows marking $\pm60\degree$ about the Z-axis. Blue and magenta dotted lines show {$\pm2.5\degree$} extensions to the larger triangle at M$_1$, M$_2$, P$_1$, and P$_2$ and {$\pm3.5\degree$ at Z$_1$ and Z$_2$. Angular deviations are exaggerated here for illustrative purpose.} Panels (c)–(e) show gamuts derived from modified MZP combinations, and (f) shows their common overlapping region {(see full text of Section~\ref{ss-gamut} for a complete discussion)}. Shaded areas outside this region indicate non-reproducible polarization measurements produced by the {foreshortening effect}.
    }}
    \label{fig:gamut}
\end{figure*} 

{A gamut in colorimetry and color reproduction refers to the range of colors that a device (like a display, printer or camera) can accurately reproduce or measure. It also describe the set of colors within a specific color space, independent of any device. A trichromatic gamut is typically visualized as a triangle in color diagrams, because the ``chroma plane'' is the plane perpendicular to the R=G=B line in the 3D color space, and a diagonal cut through a cube is a triangle. Consistent with this representation,}
\citet{DeForest_etal2022} discussed how a triplet (i.e., MZP) measurement of polarization is directly analogous to the red-green-blue (RGB) colorimetry, which \citet{Patel2023RNAAS} demonstrated using real-world polarization observations of the 2023 total solar eclipse. {However, it has been established in colorimetry that none of the standard color representation systems fully encompass all the colors visible to the human eye; instead, each system is typically limited to a specific color gamut (e.g., sRGB or Adobe RGB). Since the {foreshortening effect} alters the perceived polarization states, we encounter deviation from the ideal MZP system, resulting in a measurement that samples only portions of the full polarization gamut.}

{Following the RGB $\leftrightarrow$ MZP analogy, Figure~\ref{fig:gamut}a depicts an MZP system as a color space. Just as colors in RGB space are formed by linear combinations of base colors, polarization measurements can also be expressed as linear combinations of the MZP bases. The colors in Figure~\ref{fig:gamut}a depict the full polarization gamut within the triangle in the diagram. Because the polarizers in the MZP system are not orthogonal to each other, in the solar corona---or in fact in any other MZP measurement---when measuring polarized intensity through a given polarizer angle (e.g., Z), there is always a contribution from the other bases (M and P) as well. Consequently, pure M, Z, or P states are never produced; instead, the measurements form combinations that must fall within the circle inscribed on the full possible gamut (Figure~\ref{fig:gamut}b); the circle is a cross-section of the familiar Poincar\'e sphere, defined by the relation $I^2=Q^2+U^2+V^2$ for the familiar Stokes parameters. \citet{DeForest_etal2022} refer to it as the ``physical gamut'' of MZP measurements. This same physical gamut is observed when displaying polarized total solar eclipse images, as demonstrated in \cite{Patel2023RNAAS}.}

{When viewed face-on, the measurement gamut appears as an equilateral triangle with vertices M, Z, and P, as shown in Figure~\ref{fig:gamut}b. The reference M, Z, P axes have been extended with dashed gray arrows on the opposite side. Angular measurements are taken counterclockwise with respect to the Z axis, following the convention used for position angle measurements from solar north. Due to the {foreshortening effect}, the effective polarizer angles deviate by approximately $\pm$2.5$\degree$ and $\pm$3.5$\degree$ from the nominal $\pm$60$\degree$ and $\pm$0$\degree$ respectively. Consequently, the M and P points shift to $\mathrm{M_1 = -62.5\degree}$, $\mathrm{M_2 = -57.5\degree}$,  $\mathrm{Z_1 = 3.5\degree}$, $\mathrm{Z_2 = -3.5\degree}$, $\mathrm{P_1 = 62.5\degree}$, and $\mathrm{P_2 = 57.5\degree}$, respectively; this changes the measurement gamut slightly, limiting the polarization states of light that can be represented by the modified MZP brightnesses, just as modifying the spectrum of primary colors in an RGB system modifies the gamut of colors that can be reproduced by that system.}

{At the center of the WFI FOV where foreshortening is zero and negligible, polarization gamut within the circular region shown in Figure~\ref{fig:gamut}b will be possible. Near the top and bottom regions of the WFI FOV, where the foreshortened angles are approximately $\pm62.5\degree$, the extent of the measurement gamut becomes limited to the region above the dashed blue line $\mathrm{M_1 P_1}$. As a result, only polarizations within the region enclosed by the intersection of the circle and the triangle $\mathrm{M_1 Z P_1}$ can be measured, as shown by the unshaded region in Figure~\ref{fig:gamut}c. Similarly, near the left and right edges of the WFI FOV (Figure~\ref{fig:imax_wfi}), where the polarizer angles are reduced to approximately $\pm$58.5$\degree$, the gamut is reduced to the intersection of the circle and the triangle $\mathrm{M_2 Z P_2}$, as illustrated in Figure~\ref{fig:gamut}d. An alternate possibility with varying MZP points over the gamut is shown in panel (e). The effective intersection based on combinations where the MZP polarizer angles are extremely foreshortened within the WFI FOV is shown as the non-shaded truncated triangular region in panel (f).
Note that the truncation of gamut range for the $\pm2.5\degree$ and $\pm3.5\degree$ deviation from the M and P and Z axes respectively shown in Figure~\ref{fig:gamut} is exaggerated to enhance its visibility in the figure. In reality, the modulation is too subtle to produce a noticeable difference in this figure. Thus, the foreshortening of the angles slightly alters the range of polarizations that can be measured or represented using polarizer triplets.  The effect is important for measurements of light that is strongly polarized; but it is not significant for WFI in particular, because the overall beam entering the instrument is from a mix of polarized and unpolarized sources and therefore only weakly polarized overall.}

\subsection{Effect of Optical Distortion}
    \label{ss-distortion}

\begin{figure*}
    \centering
    \includegraphics[width=0.75\linewidth]{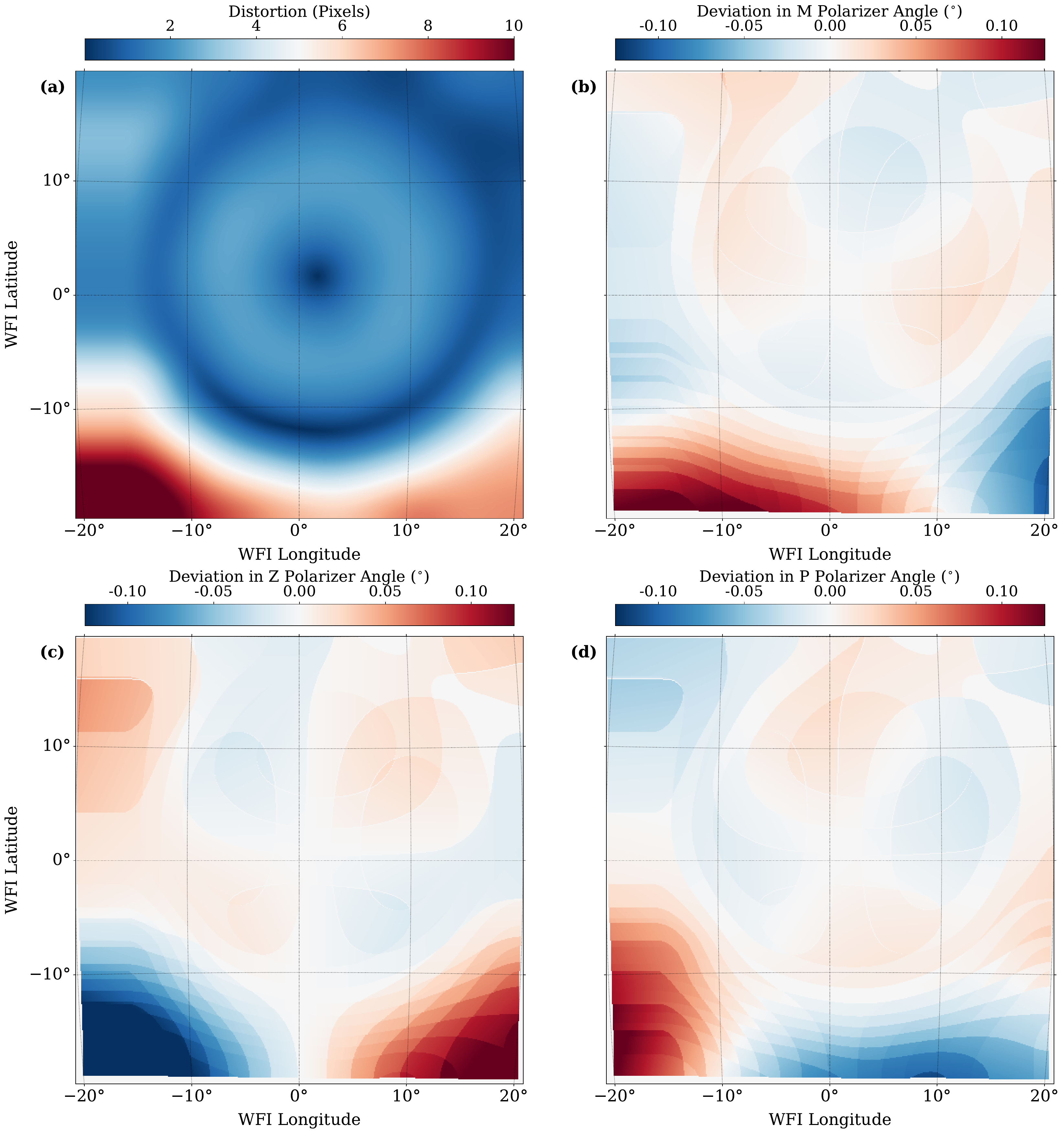}
    \caption{Effect of optical distortion on IMAX maps. (a) The distortion map of magnitude of shift in pixels ($\sqrt{x^2 + y^2}$). Panels (b)-(d) show the offset in degrees at each pixel for M, Z and P polarizer angles from Figure~\ref{fig:imax_wfi} after an optical distortion has been applied in WFI FOV.}
    \label{fig:distort}
\end{figure*}

The optical system of telescopes can introduce image distortion, for example, barrel distortion, pincushion distortion, or mustache distortion. This is especially prominent in wide field imagers like PUNCH. This distortion must be accounted in calculations across the wide FOV to determine the offset in the pixels where the light is incident on the detector. 

To measure the effects of optical distortion across the WFI FOVs, the PUNCH development team collected 540 images of the night sky at the McDonald Observatory in 2022 June with the WFI engineering model, as described in \citet{hughes2023coma}.
To characterize the optical performance of the WFI instrument, for each image we iteratively alternate between refining the pointing and distortion model. 

{Using the \texttt{punchbowl} package \citep{hughes_punchbowl_2025}, we measured the image alignment and optical distortion. The code} refines image alignment by iteratively correcting distortion. It starts with point spread-function corrected images and identifies stellar centers using Source Extraction and Photometry software \citep{Barbary2016, bertin1996}. Using non-linear least squares minimization, it updates the rotation angle and center coordinates, then builds a point-based distortion model. The location of stars in images are determined and compared with the known location of stars from the catalogs to determine the shift in horizontal ($x$) and vertical ($y$) pixels thus generating a distortion model \citep{hughes_punchbowl_2025}.


The distortion map with shift in x and y direction is displayed as the magnitude ($\sqrt{x^2 + y^2}$) and shown in Figure~\ref{fig:distort}(a). The non-symmetric pattern seen in this panel shows how the incident light is shifted in pixels due to the optical system. This distortion is then folded in with the {foreshortening effect} to result in a total effect. Panels (b)-(d) of Figure~\ref{fig:distort} show maps of pixel by pixel difference in M, Z and P polarizer angles shown in Figure~\ref{fig:imax_wfi} and the shifted angles after application of distortion at the same location. The optical distortion shifts the incident light to different pixels and hence it is required to know such small yet important deviations for improved accuracy of polarimetric measurements.
 
\section{Summary and Discussion}
    \label{sec-conclusion}

To precisely measure the polarization properties of the solar atmosphere and heliosphere, it is important to identify the geometrical effects present in the remote sensing instrument that measure it. Wide-field imagers such as PUNCH/WFI, equipped with polarizers oriented at -60$\degree$, 0$\degree$ and +60$\degree$, are designed to measure the polarimetric properties of the heliosphere with a field of view extending 40$\degree$ away from the Sun. As the signal at such distances is very faint, it is necessary to understand the effects of wide FOV on the optical system of WFI. We identified the foreshortening of the polarizer angle due to the large FOV of WFI and the geometry of the observation plane with respect to the distant target. 

We took inspiration from {movie format of} IMAX 3D presentations to account for the geometrical effects of a large screen (or FOV) in foreshortening of polarizer angles. Ground-based wide viewing angle optical systems such as wide LCD screen, monitors resolve similar effects in wide viewing angles with the aid of suitable materials along various layers in their screens. However, space-based missions have restrictions from design, space qualification, mass budget and power consumption points of view. Therefore, we aim to correct for the effects introduced in the system by post-processing the data when available on ground.

Using simple geometrical approximations, we investigated how different regions in the WFI FOV will perceive the polarizer angle based on the location of the source in the sky. We found that polarizers at angles of $\pm60\degree$ and $0\degree$ will have variations of {up to} $\pm2.5\degree$ and $\pm3.5\degree$  in WFI FOVs. This effect is negligible in field close to the center of observation and increases on moving farther from the center. {This geometric deviation is correctable in post processing, using the effective polarizer angle when interpreting the MZP triplet brightness values. If left uncorrected, it would cause small deviations in polarimetric measurements as described by \citet{DeForest_etal2022}, potentially affecting line of sight measurements of CME distance and other coronal structures \citep{Davila1994ApJ, DeForest2017ApJ}.}

Given the analogous behavior of the RGB colorspace and MZP polarizer system, an ideal instrument would cover the complete physical gamut within the MZP system, as shown in Figure~\ref{fig:gamut}a, b. However, due to the impact of the {foreshortening effect}, the measurable range of polarizations is slightly reduced depending on the location in the sky within 
the WFI FOV. This is not anticipated to affect the WFI measurements, because the light
measured by the instrument is not fully polarized and can be represented within the 
reduced measurement gamut. Left uncorrected, the variable effective polarizer angles would
form a systematic error term in the instrument's reported values for polarization states
in any system, including $pB$.  However, the effect is correctable mathematically to yield idealized MZP measurements (and corresponding $pB$ values) at each pixel.  

We also analyzed the impact of optical distortion on the {foreshortening effect}, which is important when obtaining the final image in the WFI cameras. Optical distortion deforms the image by shifting the incident signal by less than 10 pixels, with most of the image experiencing only a few pixels of distortion, while the edges are the most affected. Proper mapping is necessary to accurately record the measurement locations. Additionally, although the deviation of the perceived polarizer angle across the WFI FOV is small, it must be corrected for precise white-light polarimetry of the heliosphere.
By using the IMAX formalism described above, combined with an accurate map of the distortion the WFI polarization triplet will be corrected. The corrected images will then be used to accurately derive polarized brightness measurements \citep[See Equation 45 in][]{DeForest_etal2022}.  PUNCH-specific corrections are included in the open-source python package,\texttt{solpolpy} \citep{Patel24, hughes_2024}. \texttt{solpolpy} is a polarization resolving software suite developed in Python to transform between varying polarization bases that has been developed for the PUNCH data reduction pipeline. The {foreshortening effect} is corrected while processing the WFI data within\texttt{solpolpy}. {The \texttt{solpolpy} package also can convert MZP to other polarization bases, including the widely known Stokes (I, Q, U) parameters.} 

This analysis is significant for measuring not only polarized signal from the heliosphere, but also to characterize the polarization of the stellar background.
WFI data are combined with Narrow Field Imager (NFI) data within the PUNCH data pipeline to create a mosaic image. Given the wide FOV of the WFI instruments, accounting for distortions is crucial, as most overlapping regions are near the edges of the WFI fields of view. These data can only be combined after individual observations have been transformed from camera frames into a common polarization reference frame aligned with solar north and after {foreshortening effect} corrections, including distortion and coordinate system corrections, have been applied.

PUNCH/WFI is the widest FOV white-light polarimeter flown in space to date. The correction due to {foreshortening effect} identified for PUNCH/WFI is important and is applicable to similar current and future polarimetric instruments designed to observe such wide FOVs, including terrestrial polarimetric cameras. 

\begin{acknowledgments}
This work is funded via the PUNCH project. PUNCH is a NASA Small Explorer mission, funded via NASA Contract No. 80GSFC18C0014. DBS is also supported by NASA Heliophysics US Investigator grant 80NSSC21K1860.
\end{acknowledgments}

\software{astropy \citep{astropy2013, astropy2018},
          matplotlib \citep{Matplotlib2007},
          astrometry \citep{Astrometry2010},
          solpolpy \citep{hughes_2024}
          }

\bibliography{references}{}
\bibliographystyle{aasjournal}

\end{document}